**State of the Profession Position Paper submitted to the Astro2010 Decadal Survey *(March 2009)***

# The Revolution in Astronomy Education: Data Science for the Masses

Kirk D. Borne (George Mason University)
Suzanne Jacoby (LSST Corporation)
Karen Carney (Adler Planetarium)
Andy Connolly (University of Washington)
Timothy Eastman (Wyle Information Systems)
M. Jordan Raddick (JHU/SDSS)
J. A. Tyson (UC Davis)
John Wallin (GMU)

Abstract:

As our capacity to study ever-expanding domains of our science has increased (including the time domain, non-electromagnetic phenomena, magnetized plasmas, and numerous sky surveys in multiple wavebands with broad spatial coverage and unprecedented depths), so have the horizons of our understanding of the Universe been similarly expanding. This expansion is coupled to the exponential data deluge from multiple sky surveys, which have grown from gigabytes into terabytes during the past decade, and will grow from terabytes into Petabytes (even hundreds of Petabytes) in the next decade. With this increased vastness of information, there is a growing gap between our awareness of that information and our understanding of it. Training the next generation in the fine art of deriving intelligent understanding from data is needed for the success of sciences, communities, projects, agencies, businesses, and economies. This is true for both specialists (scientists) and non-specialists (everyone else: the public, educators and students, workforce). Specialists must learn and apply new data science research techniques in order to advance our understanding of the Universe. Non-specialists require information literacy skills as productive members of the 21$^{st}$ century workforce, integrating foundational skills for lifelong learning in a world increasingly dominated by data. We address the impact of the emerging discipline of data science on astronomy education within two contexts: formal education and lifelong learners.

## Preamble

Got data? Of course you do, and there is much more on the way! The exponential growth of data volumes in astronomy is offering new opportunities for actively involving large numbers of people in the excitement of discovery as well as posing challenges to our profession to effectively bridge the gap from data to knowledge. Since earliest human history, astronomers have been engaged in the study of dramatic and dynamic phenomena, including supernovae, eclipses, planetary motions, and more. As our capacity to study ever-expanding domains of our science has increased (including the time domain, non-electromagnetic phenomena, magnetized plasmas, and numerous sky surveys in multiple wavebands with broad spatial coverage and unprecedented depths), so have the horizons of our understanding of the Universe been similarly expanding. This expansion is coupled to the exponential data deluge from multiple sky surveys, which have grown from gigabytes into terabytes during the past decade, and will grow from terabytes into Petabytes (even hundreds of Petabytes) in the next decade. With this increased vastness of information, there is a growing gap between our awareness of that information and our understanding of it. Training the next generation in the fine art of deriving intelligent understanding from data is needed for the success of sciences, communities, projects, agencies, businesses, and economies. This is true for both specialists (scientists) and non-specialists (everyone else: the public, educators and students, workforce). Specialists must learn and apply new data science research techniques in order to advance further our understanding of the Universe. Non-specialists require information literacy skills as productive members of the 21st century workforce, integrating foundational skills for lifelong learning in a world increasingly dominated by data. We address the impact of the emerging discipline of data science on astronomy education within two contexts: formal education and lifelong learners. Recommendations are presented for:

- Training the next generation of specialists to realize the full potential of cyber-enabled science;
- Engaging students in authentic learning experiences through the use of astronomical data in secondary and undergraduate classrooms; and
- Actively involving the public in the exploration and discovery of our dynamic Universe through Citizen Science research opportunities.

As we move toward the national goal of scientific literacy for all, computational literacy must be considered a core component of that goal, with data science as an essential competency.

## The Revolution in Astronomy and Other Sciences

The development of models to describe and understand scientific phenomena has historically proceeded at a pace driven by new data. The more we know, the more we are driven to enhance or to change our models, thereby advancing scientific understanding. This data-driven modeling and discovery linkage has entered a new paradigm [1], as illustrated in the accompanying graphic [2]. The emerging confluence of new technologies and approaches to science has produced a new Data-Sensor-Computing-Model synergism. This has been driven by numerous developments, including the information explosion, the development of dynamic intelligent sensor networks [http://www.thinkingtelescopes.lanl.gov/], the acceleration in high performance computing (HPC) power, and advances in algorithms, models, and theories. Among these, the most extreme is the growth in new data. The acquisition of data in all scientific disciplines is rapidly accelerating and causing a nearly insurmountable data avalanche [3]. Computing power doubles every 18 months (Moore's Law), corresponding to a factor of 100 in ten years. The I/O bandwidth (into and out of our systems, including data systems) increases by 10% each year – a factor 3 in ten years. By comparison, data volumes appear to double every year (a factor of 1,000 in ten years). Consequently, as growth in data volume accelerates, especially in the natural sciences (where funding certainly does not grow commensurate with data volumes), we will fall further and further behind in our ability to access, analyze, assimilate, and assemble knowledge from our data collections – unless we develop and apply increasingly more powerful algorithms, methodologies, and approaches. This requires a new generation of scientists and technologists trained in the discipline of data science [4].

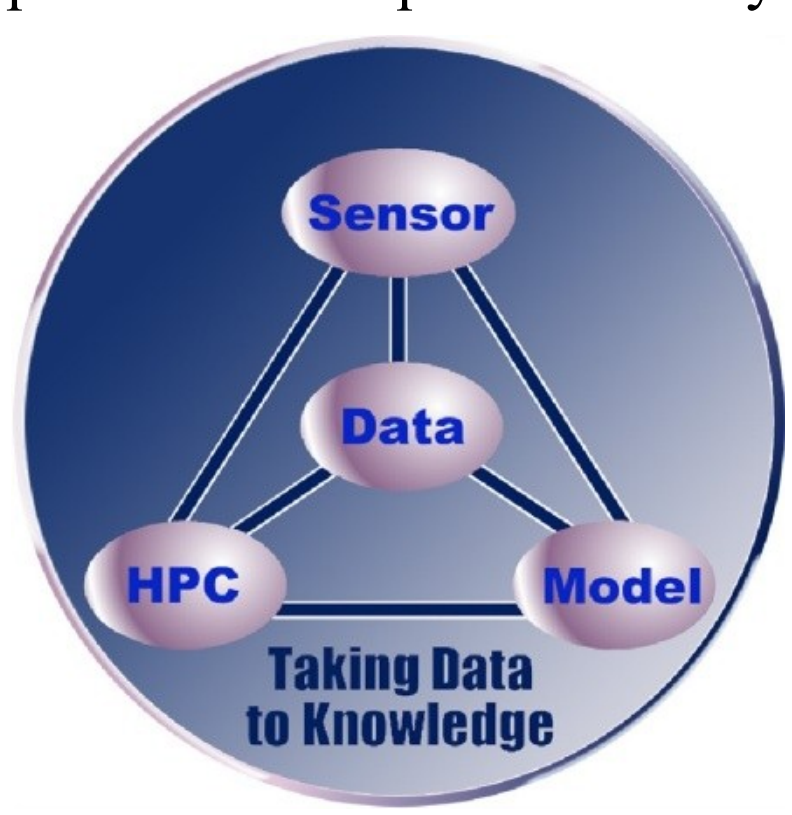

In astronomy in particular, rapid advances in three technology areas (telescopes, detectors, and computation) have continued unabated [5], all leading to more data [6]. With this accelerating advance in data generation capabilities over the coming years, we will require an increasingly skilled workforce in the areas of computational and data sciences in order to confront these challenges. Such skills are more critical than ever since modern science, which has always been data-driven, will become even more data-intensive in the coming decade [6, 7]. Increasingly sophisticated computational and data science approaches will be required to discover the wealth of new scientific knowledge hidden within these new massive scientific data collections [8, 9].

The growth of data volumes in nearly all scientific disciplines, business sectors, and federal agencies is reaching historic proportions. It has been said that *"while data doubles every year, useful information seems to be decreasing"* [10], and *"there is a growing gap between the generation of data and our understanding of it"* [11]. In an information society with an increasingly knowledge-based economy, it is imperative that the workforce of today and especially tomorrow be equipped to understand data and to apply methods for effective data usage. Required understandings include knowing how to access, retrieve, interpret, analyze, mine, and integrate data from disparate sources. In the sciences, the scale of data-capturing capabilities grows at least as fast as the underlying microprocessor-based measurement system [12]. For example, in astronomy, the fast growth in CCD detector size and sensitivity has seen the average dataset size of a typical large astronomy sky survey project grow from hundreds of gigabytes 10 years ago (e.g., the MACHO survey), to tens of terabytes today (e.g., 2MASS and Sloan Digital Sky Survey [5]), up to a projected size of tens of petabytes 10 years from now (e.g., LSST, the Large Synoptic Survey Telescope [6]). In survey astronomy, LSST will produce one 56Kx56K (3-Gigapixel) image of the sky every 20 seconds, generating nearly 30 TB of data daily for 10 years. In solar physics, NASA announced in 2008 a science data center specifically for the Solar Dynamics Observatory, which will obtain one 4Kx4K image every 10 seconds, generating one TB of data per day. NASA recognizes that previous approaches to scientific data management and analysis will simply not work. We see the data flood in all sciences (e.g., numerical simulations, high-energy physics, bioinformatics, geosciences, climate monitoring and modeling) and outside of the sciences (e.g., banking, healthcare, homeland security, drug discovery, medical research, retail marketing, e-mail). The application of data mining, knowledge discovery, and e-discovery tools to these growing data repositories is essential to the success of our social, financial, medical, government, and scientific enterprises. **An informatics approach is required.** What is informatics? Informatics has recently been defined as *"the use of digital data, information, and related services for research and knowledge generation"* [13], which complements the usual definition: *informatics is the discipline of organizing, accessing, integrating, and mining data from multiple sources for discovery and decision support* [14].

## A National Imperative

Our science education programs have always included the principles of evidence-based reasoning, fact-based induction, and data-oriented science [15]. In this age of the data flood, greater emphasis on and enhancement of such data science competencies is now imperative. In particular, we must muster educational resources to train a skilled data-savvy workforce: one that knows how to find facts (i.e., data, or evidence), access them, assess them, organize them, synthesize them, look at them critically, mine them, and analyze them.

The *Nature* article *"Agencies Join Forces to Share Data"* calls for more training in data skills [16]. This article describes a new Interagency Working Group on Digital Data representing 22 federal agencies in the U.S., including the NSF, NASA, DOE, and more. The group plans to set up a robust public infrastructure so that all researchers have a permanent home for their data. One option is to create a national network of online data repositories funded by the government and staffed by dedicated computing and data science professionals with science discipline expertise. Who will these computing and archiving professionals be? They will be a professional workforce trained in the disciplines of computational and data sciences and who collaborate with computer science and statistics professionals in these areas, including machine learning, visualization, statistics, algorithm design, efficient data structures, scalable architectures, effective programming techniques, information retrieval methods, and data query languages.

Within the scientific domain, data science is becoming a recognized academic discipline. F. J. Smith argues that now is the time for data science curricula in undergraduate education [17]. Others promote data science as a rigorous academic discipline [18]. Another states that *"without the productivity of new disciplines based on data, we cannot solve important problems of the world"* [19]. The 2007 NSF workshop on data repositories included a track on data-centric scholarship – the workshop report explicitly states our key message: *"Data-driven science is becoming a new scientific paradigm – ranking with theory, experimentation, and computational science"* [20]. Consequently, astronomy and other scientific disciplines are developing sub-disciplines that are information-rich and data-intensive to such an extent that these are now becoming (or have already become) recognized stand-alone research disciplines and full-fledged academic programs on their own merits. The latter include bioinformatics and geoinformatics, but will soon include astroinformatics, health informatics, and data science.

## National Study Groups Face the Data Flood

Several national study groups have issued reports on the urgency of establishing scientific and educational programs to face the data flood challenges:

1. NAS report: *"Bits of Power: Issues in Global Access to Scientific Data"* (1997) [21];
2. NSF report: *"Knowledge Lost in Information: Report of the NSF Workshop on Research Directions for Digital Libraries"* (2003) [22];
3. NSB (National Science Board) report: *"Long-lived Digital Data Collections: Enabling Research and Education in the 21st Century"* (2005);
4. NSF report with the Computing Research Association: *"Cyberinfrastructure for Education and Learning for the Future: A Vision and Research Agenda"* (2005);

5. NSF “Atkins Report” : *“Revolutionizing Science and Engineering Through Cyberinfrastructure: Report of the National Science Foundation Blue-Ribbon Advisory Panel on Cyberinfrastructure”* (2005) [23];
6. NSF report: *“The Role of Academic Libraries in the Digital Data Universe”* (2006) [24];
7. NSF report: *“Cyberinfrastructure Vision for 21st Century Discovery”* (2007) [25];
8. JISC/NSF Workshop on Data-Driven Science & Repositories (2007) [20].

Each of these reports has issued a call to action in response to the data avalanche in science, engineering, and the global scholarly environment. For example, the NAS *“Bits of Power”* report lists five major recommendations, one of which includes: *“Improve science education in the area of scientific data management”* [21]. The Atkins NSF Report stated that skills in digital libraries, metadata standards, digital classification, and data mining are critical [23]. In particular, that report states: *“The importance of data in science and engineering continues on a path of exponential growth; some even assert that the leading science driver of high-end computing will soon be data rather than processing cycles. Thus it is crucial to provide major new resources for handling and understanding data.”* [23] The core and most basic resource is the human expert, trained in key data science skills. As stated in the 2003 NSF “Knowledge Lost in Information” report, human cognition and human capabilities are fundamental to successful leveraging of cyberinfrastructure, digital libraries, and national data resources [22].

**Cyberinfrastructure, Human Computation, and Computational Thinking**

New modes of discovery have been enabled by the growth of computational infrastructure in the sciences. This cyberinfrastructure includes massive databases, virtual observatories (distributed data), high-performance computing (clusters and petascale machines), distributed computing (the Grid, the cloud, and peer-to-peer networks), intelligent search and discovery algorithms, innovative visualization methods, collaborative research environments, Web 2.0 tools, and more. To cope with the data flood, a paradigm shift is required – we must go beyond cyber-enabled discovery to human-powered discovery. By that we mean “human computation” [26] and “computational thinking” [27]. Human computation refers to *“Tasks like image recognition that are trivial for humans, but which continue to challenge even the most sophisticated computer programs”* [26]. Computational thinking refers to a new kind of literacy, akin to math or cultural literacy. It is an epistemological orientation (a way of thinking and knowing) that is consistent with computational methods of organizing data and processes. Computational thinking addresses the paradox of machine intelligence: knowing which tasks are best assigned to computers, and which are best assigned to humans. A writer addressed the application of these concepts to the astronomical data flood by stating the absolute necessity of public involvement with astronomical data archives: ‘*Though these "virtual observatories" are used primarily by professionals, they can also be welcoming to educators, students and amateur astronomers, who get instant access to the best telescopes in the world. And why not open the doors wide? It's hard to imagine that this data will ever get "used up" – that all the good discoveries will one day be wrung out of it – so the more minds working away at it, the better.’* [28] Of course, human factors research related to human-computer interaction is absolutely essential for this to succeed.

**Target Audiences**

We address the impact of these data science issues on astronomy education within two contexts: Formal Education, including K-12 and undergraduate STEM (Science, Technology, Engineering, and Mathematics) courses, and life-long learners. The latter include citizen scientists, who are *trained volunteers who work on authentic science projects with scientific researchers to answer real-world questions*. It is well known that among the sciences, two topics seem to grab the most public devotion and attention: dinosaurs and Space! For example, the release of Google Sky generated millions of users within the first few days, and the announcement of Galaxy Zoo (described later) attracted 80,000 volunteers within a few weeks. Much of the reason for this is intangible: emotion and affective motivations [29].

**Modes of Interaction**

We consider large sky surveys, with their open data policies ("data for everyone") and their uniformly calibrated massive databases, as key cyberinfrastructure and the major content providers in the education of astronomical data scientists and in astronomy-based data science curricula. For different audiences and in different learning settings, we envision three broad modes of interaction with sky survey data (i.e., database catalogs and image archives), which represent a progression from information-gathering to active engagement to discovery:

a) Data Discovery – What was observed, when, and by whom? Retrieve observation parameters from an astronomical sky survey catalog database. Retrieve parameters for interesting objects.
b) Data Browse – Retrieve images from a sky survey image archive. View thumbnails. Select data format (JPEG, Google Sky KML, FITS). Pan the sky and examine catalog-provided tags (Google Sky, World Wide Telescope).
c) Data Immersion – Perform data analysis, mining, and visualization (via unified software tools for astronomy education). Report discoveries. Comment on observations. Contribute followup observations. Engage in social networking, annotation, and tagging. Provide classifications of complex images, data correlations, data clusters, or novel (outlying) detections.

**Formal Education: K-12 and Undergraduate**

Astronomy provides an innately engaging scientific context within which teachers can engage students in research investigations that make use of publicly accessible databases. These engaging experiences support both the learning of science and the development of 21st century workforce skills. Astronomical sky survey data can become a key part of projects emphasizing student-centered research in middle school, high school, and undergraduate settings. Professional development, including the preparation and retention of highly qualified teachers, plays a critical role. The importance of teachers cannot be underestimated. The most direct route to improving mathematics and science achievement for all students is better mathematics and science teaching [30]. In fact, *"teacher effectiveness is the single biggest factor influencing gains in achievement, an influence bigger than race, poverty, parent's education, or any of the other factors that are often thought to doom children to failure.*" [31] The next-generation astronomy-based STEM education program will involve student research projects in conjunction with teacher professional development programs. Taught in an exemplary fashion, using technology to its best advantage, students can participate in cutting-edge discovery with authentic classroom research opportunities developed through such efforts.

As described in "How People Learn" [32], the goal of education is to help students develop needed intellectual tools and learning strategies, including how to frame and ask meaningful questions about various subject areas. This ability contributes to individuals becoming self-sustaining, lifelong learners. Engaging students by using real data to address scientific questions in formal education settings is known to be an effective instructional approach toward this goal.

Specifically, the *National Science Education Standards* [33] emphasize that students should learn science through inquiry (Science Content Standard A: Science as Inquiry) and should understand the concepts and processes that shape our natural world (Science Content Standard D: Earth and Space Science). Students learn best if they are not passive recipients of factual information but rather are engaged in the learning process. The Socratic method tells us that *"teaching is not telling; students must be involved in more than listening to learn."*

One goal of having teachers and their students engage in data analysis and data mining is to help them develop a sense of the methods scientists employ, as well as a familiarity with the tools they use to "do science." This is a critical component in understanding the process of science, including procedures for data collection, analysis of bias, and scientific interpretation. The common lecture-textbook-recitation method of teaching, still prevalent in today's high schools, tends to discourage students from applying important scientific, mathematical, and technological skills in a meaningful context. Robert Yager states that this model of teaching science is akin to teaching all the rules of a sport, like softball, to a child—how to bat, catch, throw, slide, and wear the uniform—but never letting the child actually play in a game! [34] Recent research has in fact shown that students benefit more from depth rather than breadth in their high school science courses [35]. One study examined over 8000 college students in biology, chemistry, and physics. We believe that the teaching of data science methodologies as a means of increasing the depth of study in core science courses will have lasting benefit for a lifetime, as well as immediate tangible benefit in a student's college education. Additional research has demonstrated the efficacy of using data in the classroom for improving student learning in science – the report "Using Data in Undergraduate Classrooms" is especially relevant, with valuable suggestions, education research results, pedagogical insights, teaching scenarios, and resource listings [36].

A particularly acute problem is in undergraduate Physics programs (presumably still the source of most astronomy Ph.D. students). These programs rarely require much in the way of computational or statistical (data) sciences training, though they may require a computer language programming course. Courses in computational science, scientific computing, numerical modeling and simulation, scientific data and databases, and scientific data mining should be developed and offered as electives, if not requirements. The new CDS (Computational and Data Sciences) B.S. degree program at GMU is one of the first to require such courses [37].

### Citizen Science: Life-Long Learners

The formal education system does not exist in a vacuum; students, teachers, and families are part of a broader context for learning. Rich opportunities for learning outside the classroom include Informal Science Education, Out of School Time (OST), and the world of Citizen Science, where non-specialist volunteers assist scientists' research efforts by collecting, organizing, or analyzing data. More than a decade of research shows that sustained participation in well-executed OST experiences can lead to increases in academic achievement and positive impact on a range of social and developmental outcomes [38]. *"Experiences in informal settings can significantly improve science learning outcomes for individuals from groups which are historically underrepresented in science, such as women and minorities. Evaluations of museum-based and after-school programs suggest that these programs may also support academic gains for children and youth in these groups."* [39] Engagement in informal education, visits to museums and planetaria, and now Citizen Science can help to create an environment that encourages young people to pursue challenging courses (and maybe careers) in STEM disciplines.

Citizen Science is one approach to engaging the public in authentic scientific research. **Citizen science** is a term used for projects or ongoing programs of scientific work in which individual volunteers or networks of volunteers, many of whom may have no specific scientific training, perform or manage research-related tasks such as observation, measurement, or computation. A recent and highly successful astronomy Citizen Science project, Galaxy Zoo, has involved more than 185,000 armchair astronomers from all over the world in classifying the morphology of galaxies from SDSS, resulting in four papers published in peer-reviewed journals (e.g., [40, 41]). In the 18 months prior to February 2009, 80 million classifications of galaxies were submitted on one million objects at galaxyzoo.org. The Stardust@home project [42], where volunteers must pass a test to qualify to participate in the search for grains of dust in aerosol gels from the NASA Stardust Mission, has attracted a smaller number of citizen scientists (24,000), perhaps because of the more sophisticated training and analysis required by participants, or perhaps because images of galaxies are inherently more interesting to the larger public than cracks in an aerosol gel. The accompanying figure illustrates Design Considerations for Citizen Science Projects showing three overlapping circles: Research the Public Can Do, Research that is Interesting to the Public, and Research Scientists Care About. Finding out exactly why a particular project occupies one portion of the chart over another is a key part of the research agenda for Citizen Science. In all cases, citizen scientists work with real data and perform duties of value to the advancement of science. The human is still better at pattern recognition (Galaxy Zoo) and novelty (outlier) detection (Stardust@home) tasks than a computer, making the galaxy classification activity and others like it good candidates for successful Citizen Science projects.

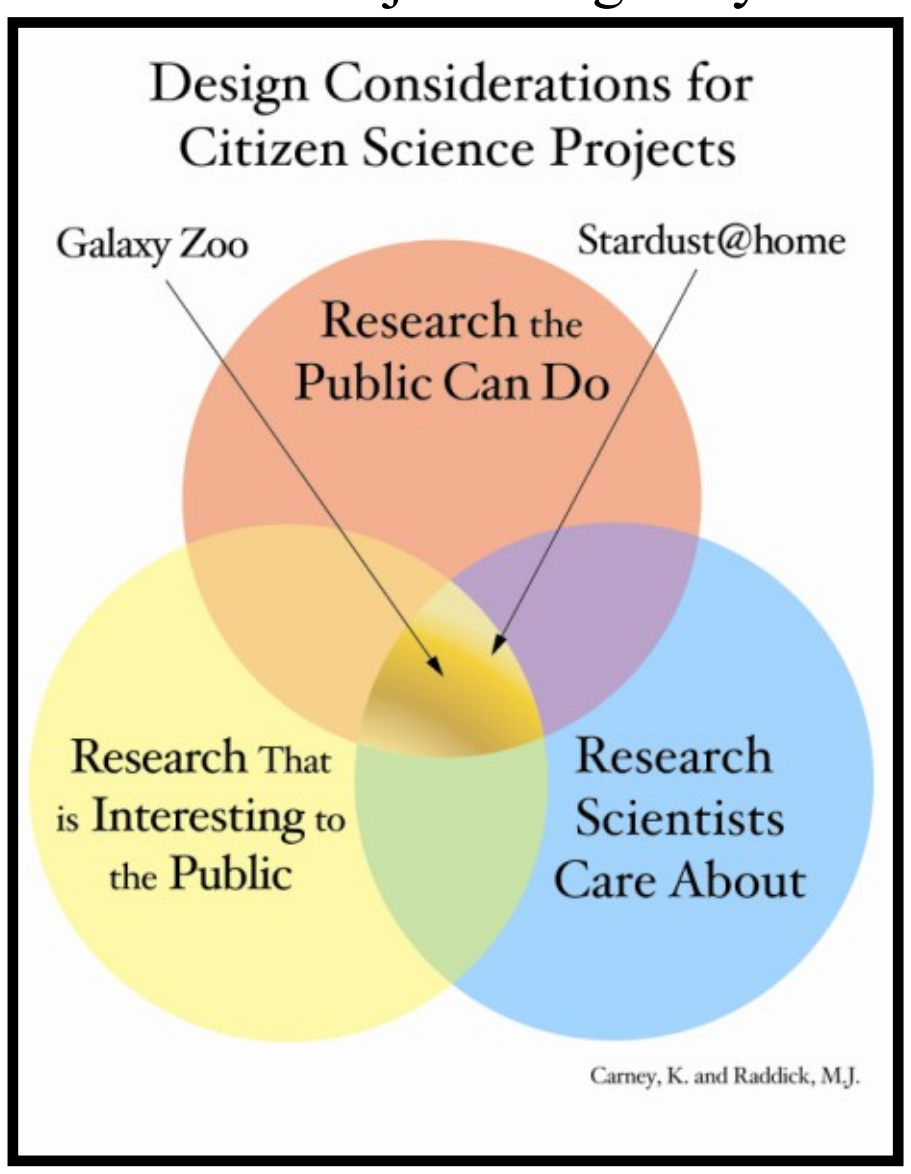

Several popular pathfinders exist for Citizen Science projects, which include: (a) passive engagement projects (e.g., SETI@home [43]) in which the participant sets up a screensaver on their computer and the rest is done automatically, with very little human contribution or expertise; and (b) active involvement activities (e.g., the AAVSO [44], and the Audubon Society Christmas Bird Count [45]) in which the participant becomes an active contributing scientist, lending their human cognitive abilities to the acquisition of knowledge from large complex natural systems. We strongly support the deployment of more of the latter experiences, which contribute significantly to lifelong learning and science education among a broader audience.

An exciting prospect for Citizen Science research projects enters with the advent of time domain astronomy. For example, starting near the middle of the next decade, the LSST will obtain one 10-square degree image of the sky every 20 seconds, will march along systematically to image the entire visible sky within a 3-night period, and will repeat this for 10 years. Within each image, LSST will identify every object that has moved, changed brightness, or (dis)appeared. LSST's key science theme is Mapping the Dynamic Universe. Within this context, predicted counts of known astronomical transients, solar system objects, supernovae, other dynamic phenomena, and extrapolations to all types of "unknown unknowns" lead to a combined estimate of 100,000 events to be detected each night by LSST. The LSST Science Collaboration teams have generated ideas for Citizen Science research projects that engage the public in monitoring, classifying, and annotating these events for the advancement of astronomical research [46].

**Recommendations**

We promote the teaching of astronomy (at any grade level and in all settings) as:

a) a forensic science – evidence-based inquiry from data collections, learning from data;
b) a dynamic science – the changing universe, time domain astronomy;
c) a platform for improving STEM education – using the appealing forensic nature of astronomy as a hook (a detective story to be solved) to draw students into learning, developing critical thinking skills, using fun real-life data experiences to learn key data science skills (visualization, analysis, synthesis, deduction, inference, interpretation; and
d) a rich source of life-long learning – in which Citizen Science and science centers draw the interested masses into the joy, excitement, and wonder of learning about our Universe.

We specifically make the following recommendations to the State of the Profession Group on Astronomy Education:

1. To high schools, colleges, and universities: Provide data-centric research experiences in STEM courses. Promote the use of real scientific data in the classroom through mini-grants, curriculum mandates, and/or a faculty reward system. Integrate education and research in science. Teach critical data science skills to all students as part of their science graduation requirements. Transform education using databases and information resources in all disciplines. Establish mentorship programs for minority and underrepresented groups to seed, support, and sustain their participation in science for a lifetime. Grow curricula (programs and courses) in data science (in general) and in astronomical data science (i.e., astroinformatics, specifically). Promote computational literacy across the curriculum. Hire faculty with data science training and research experience.

2. To national labs, research centers, and science data archive facilities: Permit science staff to spend a few percent of their time in Citizen Science outreach projects. Provide professional development opportunities for educators and their students to work in your research groups using real scientific data. Develop resource centers for publicly accessible and usable science data products and services. Create digital libraries via educator forums that archive specifically data science curricula materials for different core science courses (including physics and astronomy) as a mechanism for easy transfer of knowledge, lesson plans, and projects between informal and formal education venues. Promote diversity in the workplace through stimulus and reward programs in the computational and data sciences.

3. To researchers and project teams: Develop citizen science projects to engage the public and to take advantage of their cognitive capabilities (human computation and computational thinking). Reach out to minority and underrepresented communities to broaden participation in science and discovery. Work with informal environments which already have knowledge about motivation and engagement in free-choice learning environments (e.g., [47]).

4. To funding agencies: Mandate open data policies for large projects; encourage this for all projects. Mandate an outreach component in all major projects and missions – reward innovative public uses of mission/project data. Through the NSF MREFC funding pipeline, support construction of infrastructure that facilitates sharing of data products with a broad public audience, not just professional researchers. Support the development and integration of emerging transformative computational technologies into the science funding pipelines. Develop programs that fund the development of curricula and educational programs at the intersection of astronomical and data sciences. Provide viable and sustained funding for: (a) cross-disciplinary programs that promote cyber-enabled discovery in the sciences; (b) data science research; (c) early career grants aimed at data-intensive science research; (d) informal education and human computation initiatives that extend the discovery potential of large science data sets (Citizen Science); (e) education research in the science of learning from large data sets; (f) outreach involving large sky surveys and large astronomy databases and archives; (g) development of unified data tools for education (including data analysis, mining, visualization, and manipulation); (h) teacher training workshops focused on "using data in the classroom"; and (i) development of astroinformatics as a stand-alone research and education sub-discipline of astronomy.

5. To professional societies: Motivate and promote the connection between data-intensive science and classroom learning. Identify the connections between astronomical science and national education standards in all disciplines (not just math and science). Support partnerships with major online providers of information (e.g., Google, Microsoft, Yahoo!) to get the data out to a wider audience. Endorse and promote career paths for professionals with data science skills, in the same manner in which instrument-builders are afforded tenure rank and professional standing in academic departments.

Our position is that public support, through Citizen Science and human computing, is essential to the continued health of the astronomy profession. We advocate increased support of the public and educational aspects of astronomy outreach, which are easily facilitated through large sky survey projects with open data policies. We promote the use of astronomical databases both as an effective mechanism in formal STEM education (through data-centric research experiences in the classroom) and as a way to teach data science skills relevant to all disciplines and citizens.

**Position Paper Contributors and Co-Signers:**

**Primary Contributors**

Kirk D. Borne (George Mason University)
Suzanne Jacoby (LSST Corporation)
Karen Carney (Adler Planetarium)
Andy Connolly (University of Washington)
Timothy Eastman (Wyle Information Systems)
M. Jordan Raddick (JHU/SDSS)
J. A. Tyson (UC Davis)
John Wallin (GMU)

**Co-Signers:**

Jacek Becla (SLAC)
Michael Castelaz (PARI)
Alanna Connors (Eureka Scientific)
Tim Hamilton (Shawnee State U.)
Chris Lintott (Oxford)
Bruce McCollum (Caltech)
Peter Fox (RPI)
Ashish Mahabal (Caltech)
Julia Olsen (U. Arizona)
Misha Pesenson (Caltech/IPAC)
Andrew Ptak (JHU)
Nic Ross (Penn State U.)
Andrea Schweitzer (Little Thompson Observatory)
Terry Teays (JHU/MD Space Grant Consortium)
Michael Way (NASA/Goddard Institute for Space Studies)
Michael Wood-Vasey (University of Pittsburgh)